# The Astronomy of the Kamilaroi People and their Neighbours

Robert S. Fuller[1,2], Ray P. Norris[1,3], Michelle Trudgett[1]

[1]Department of Indigenous Studies, Warawara, Macquarie University NSW 2109, Australia
Email: *robert.fuller1@students.mq.edu.au, michelle.trudgett@mq.edu.au*
[2]Macquarie University Research Centre for Astronomy, Astrophysics and Astrophotonics,
Macquarie University NSW 2109, Australia
[3]CSIRO Astronomy and Space Science, PO Box 76, Epping, NSW, 1710, Australia
Email: *raypnorris@gmail.com*

## Abstract

The Kamilaroi people and their neighbours, the Euahlayi, Ngemba, and Murrawarri, are an Aboriginal cultural grouping located in the northwest and north central of New South Wales. They have a rich history, but have been missed in much of the literature concerned with sky knowledge in culture. This study collected stories, some of which have not previously been reported in an academic format, from Aboriginal people practicing their culture, augmented with stories from the literature, and analysed the data to create a database of sky knowledge that will be added to the larger body of Aboriginal cultural knowledge in Australia. We found that there is a strong sky culture reflected in the stories, and we also explored the stories for evidence of an ethnoscientific approach to knowledge of the sky.

## Notice to Aboriginal and Torres Strait Islander Readers

This paper contains the names of people who have passed away, and refers to events, such as massacres, which may be upsetting.

## 1. Introduction

The Kamilaroi people have a long history in Australia. Aboriginal Australians are descendants of people who left the Middle East approximately 70,000 years ago (Rasmussen et al. 2011: 98). Archaeological evidence of the arrival of Aboriginal people in Australia provides a wide range of dates of settlement. Archaeology has dated the Mungo Man burial in the Willandra Lakes region of New South Wales (NSW) to more than 40,000 BP (Bowler et al, 2003: 840). This supports evidence of a "fast-track" spread across the Pleistocene continent of Sahul (Australia, New Guinea, and Tasmania). Therefore, Aboriginal people may have lived in the area included in this study, north and northwest NSW, for at least 40,000 years.

When Major Thomas Mitchell, the government explorer, made the first expedition into this region in 1831, he found bushrangers and drovers settling in the area between the Liverpool Range and what is now Tamworth; a rich agricultural area now known as the Liverpool Plains. As the first writer about this country, Mitchell's comments on the "native" population provide a limited insight into the culture of the Aboriginal peoples of this area before further European settlement. Mitchell indicated that the people lived in a rich environment, with many resources around the rivers that they utilised (Mitchell, 1838: 84, 95, 100). His few remarks on language include *Einèr* for Aboriginal woman, which is remarkably similar to the Kamilaroi *Yinarr*, and appears to show contact with Kamilaroi people (ibid: 78; Ash et al. 2003: 155). In the period after Mitchell's first expedition, pressure from European settlers, particularly in the Liverpool Plains region, led to resistance from the Kamilaroi people in the area, and a resultant rapid increase in conflict. Broome (1988: 101) estimates that 16 European, and 500 Aboriginal people died between 1832 and 1838 in the region, many from historically known massacres of Aboriginal people.

Sveiby and Skuthorpe (2006: 25) have estimated the population of the Kamilaroi cultural group before settlement to be 15,000. By 1842, it is thought that there were approximately one thousand Kamilaroi in the area (ibid: 26). The rapid reduction of population from 15,000 to 1,000 in the 54 years after European settlement resulted in a displacement of the Aboriginal people of this cultural bloc towards the northwest of the Kamilaroi area (confirmed by participant P2). The current population of people identifying as having Kamilaroi ancestry is approximately 26,000, and Euahlayi ancestry, 3,000 (estimates from Kamilaroi Nation Applicant Board provided by M. Anderson).

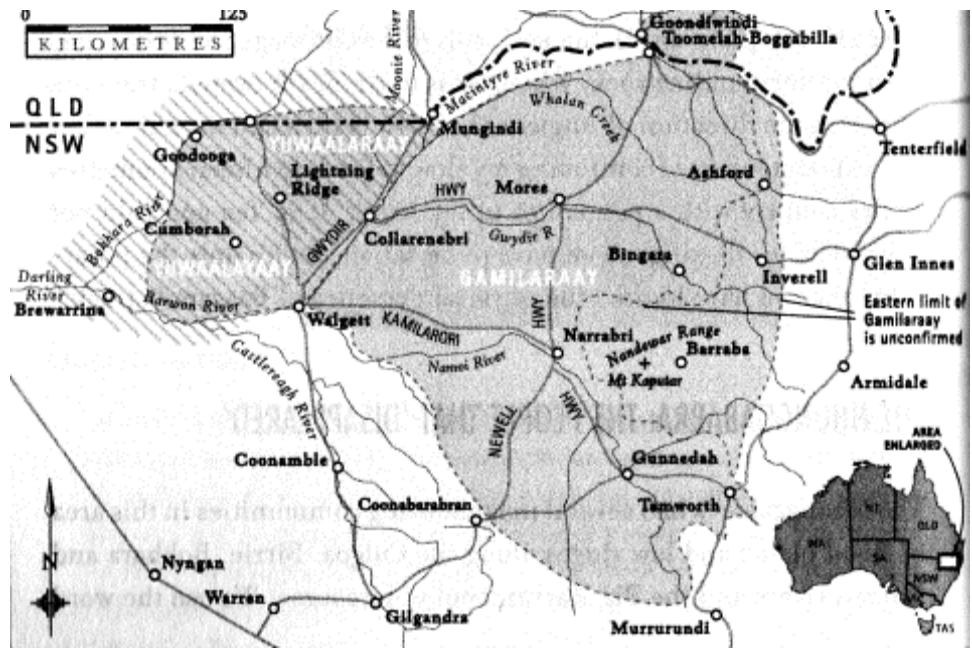

*Fig 1 the approximate area of the cultural group and languages spoken (Sveiby & Skuthorpe, 2006)*

The geographic boundaries of the area defined for this study depended very much on the definition of which Aboriginal language groups were included in this study. Mathews (1900: 576, Plate VIII) showed an area of northern NSW roughly encompassing the modern understanding of Kamilaroi country, but including the Darling River, thus including a number of smaller language groups to the West. Howitt (1904: 50) defined the Kamilaroi area as: "In short, nearly the whole of the pastoral district of Liverpool Plains." The only Aboriginal description of the area (Sveiby and Skuthorpe; 2006: 24) described a cultural group of 26 communities (including the Kamilaroi) as "bounded by the Barwon River to the east, the Warrego River to west, and the Bogan River to the south" (in their map, shown in Fig. 1, the northern boundary approximates the NSW/Queensland [QLD] border). We used this definition for the boundaries of this study, and added a limited survey of the Murrawarri and Ngemba language groups, who are located to the West and southwest of the Kamilaroi/Euahlayi community (one of the participants, P2, said that Sveiby and Skuthorpe were incorrect in using the Warrego River as the west boundary of the cultural area, and they should have used the Culgoa River).

## 1.2. Recording the culture of the Kamilaroi

Ethnology, the study of the cultures of man, was formalised by E.B. Tylor in 1865 (Radin, 1929: 9). The literature about Aboriginal culture during the period 1850-1900 changes from that written by country police and members of the clergy, to that of the first ethnologists and then anthropologists.

A major factor in examining the published information from the 19[th] century is that there were over 250 separate language groups existing at the time of European settlement (Walsh, 1991: 27). Many language groups barely survived long enough after European settlement to appear in the anthropological record as a distinct group.

There are only a limited number of early sources on Kamilaroi culture, often from the northwest of the Kamilaroi area, including Ridley (1856, 1872, 1875, 1878), Fraser (1888), Greenway (1878, 1901), and Fison & Howitt (1880). R.H. Mathews was a surveyor who worked in the area extensively, eventually publishing a large body of work (Matthews 1900, 1904, 1905) about Aboriginal religion, folklore, and ceremony, with a strong emphasis on southeast Australia. Contemporary with Mathews was K. Langloh Parker, the wife of a grazier on the Narran River, who formed a close relationship with the local Euahlayi people, and collected a large body of folklore (Parker, 1898, 1914; Parker & Lang 1897, 1905). The only recent attempt to describe Kamilaroi culture is *"Treading Lightly"* by Sveiby and Skuthorpe (2006), which is mainly about the Nhunggabarra band of the Euahlayi language group.

## 1.3. Archaeoastronomy and cultural astronomy

Archaeoastronomy is the interdisciplinary study of ancient, prehistoric, and traditional astronomy and its cultural context (Krupp, 1994: ix). Cultural astronomy is the study of the effect of astronomical knowledge or theories on ideologies or human behaviour. (Campion, 2003: xv).

In the Australian Aboriginal context, archaeoastronomy and cultural astronomy were first reported by Stanbridge (1857), who wrote about the sky knowledge of the Boorong people (Hamacher and Frew, 2010). Subsequent writers have reported Aboriginal knowledge and stories of the sky, but it wasn't until Mountford (1956: 479-504) wrote extensively about this aspect of the 1948 Arnhem Land expedition that it was treated as a separate subject. Mountford (1976: 449-483), similarly wrote about this aspect of central desert culture and then Haynes (1996) wrote extensively on the subject, and suggested that Australian Aboriginal people were "the world's first astronomers" (ibid: 7). Subsequently, Johnson (1998) has written a study of general cultural astronomy knowledge from a wide selection of language groups, and Cairns & Harney (2003) have written a detailed study of the Wardaman people's knowledge. A number of other researchers, including Clarke (2007), Hamacher and Norris (2009, 2010, 2011), Norris (2007), Norris and Hamacher (2009, 2011), Norris and Norris (2009), and Tindale (1983), have developed the current body of knowledge of Aboriginal cultural astronomy, both in general, and for specific language groups. This body of knowledge is now growing rapidly, and includes evidence that Aboriginal stone arrangements and bora sites included links to the sky (Fuller et al, 2013; Hamacher et al, 2013; Norris et al, 2013).

The use of sky knowledge in prediction of resources, and in cultural interpretation of law, has long been accepted, but we also describe the use of astronomy in an ethnoscientific manner. We define ethnoscience as an intellectual endeavor to describe the natural world within an appropriate cultural context, resulting in predictive power and practical applications such as navigation or timekeeping. Ethnoscience therefore has the same intellectual goals as modern-day western science, but without the trappings associated with Western culture and scientific tradition. For example, the Yolngu people explained the tides as the Moon filling and emptying as it passed through the ocean at the horizon, which is a perfectly sound explanation given the available evidence. This evidence-based approach to understanding the world in an appropriate cultural context correctly predicted how the height of the tide varies with the phase of the Moon, with the highest tide ("spring tide") occurring at the full or new moon (Norris and Hamacher, 2011: 4).

While there is a growing body of knowledge of Aboriginal cultural astronomy, there has been no corresponding study of the Kamilaroi language group and their neighbours, and this study was developed to fill the gap through a comprehensive survey and analysis of the literature and current knowledge.

## 1.4. Expectation of finding knowledge of Kamilaroi cultural astronomy

Comprehensive studies of the literature on Aboriginal astronomy (Fredrick, 2008; Hamacher, 2011) have shown that there is a rich knowledge of Aboriginal astronomy in the literature, much of which is based on ethnographic work. This knowledge varies from simple vocabularies up to complete cosmologies based on the relationship between the Earth and the sky. One of the side effects of this knowledge is a controversy best described by Swain (1993) as a breaking of the relationship between Aboriginal culture and place through the effects of invasion and colonisation, to be replaced by a utopian return to a sky world on death, rather than return to place. As pre-colonisation Aboriginal culture was predominantly oral in nature, there will never be a "satisfactory reconstruction of pre-colonial views of the world in southeast Australia" (ibid: 119). This could mean that all knowledge in the literature is post-colonial, and could be tainted with the influence of changes brought about by the process. One of the questions studied here is whether there is any evidence that the knowledge collected had a pre-invasion origin. Discussions with participants in the research for this paper did reveal strongly held beliefs that, contrary to Swain (1993), the sky always formed part of the connection between place and the person in Aboriginal culture.

Literature on the Kamilaroi culture includes references to knowledge of the sky. As early as 1856, Ridley's vocabulary included words in Kamilaroi for "star" and "Sun" (Ridley, 1856: 290), and a later work included the words for "Venus" and many other objects (Ridley, 1875: 26). The early ethnographic literature about the Kamilaroi's neighbours, the Euahlayi, Murrawarri, and Ngemba, also contained references to knowledge of the sky. For example, Parker and Lang (1897, 1898, 1905, 1914) gives detailed references to the sky, including the Sun, Moon, Pleiades, and meteors.

In this paper, we collected knowledge of the sky from the Kamilaroi, Euahlayi, Ngemba, and Murrawarri communities, both from the literature and from participants. We then used this knowledge to test:

- Hypothesis 1: that knowledge from these language groups could add to the current body of knowledge of Aboriginal sky culture, and;

- Hypothesis 2: that the Kamilaroi and their neighbours had an ethnoscientific knowledge of the night sky through observation and experimentation.

## 2. Methodology

### 2.1 Literature review

Using modern search engines, and large databases of publications, subject-specific literature research has a high probability of finding the majority of existing documents relevant to the subject studied. It was expected that there would be a limited number of references on sky knowledge of the Kamilaroi and their neighbours, but a large number of references on culture which would have to be searched for information of interest to this study. Some resources included in the review of literature were the State Library of NSW, the National Library, and the AIATSIS library.

### 2.2 Ethnographic research

If the Kamilaroi people were pushed out of their traditional country and suffered serious loss of culture, then there may have been no large-scale ethnographic research done on the Kamilaroi in the $20^{th}$ century, other than the extensive work of Janet Mathews (Thomas, 2011: 285, 344) in the 1960's and 70's, mainly on language. The literature review would confirm this, but in any case, the question remained; would there be any value in ethnographic research among the Kamilaroi today? This project was undertaken as a suggestion by one of the co-authors (Norris) who had been contacted by a Kamilaroi man (Greg Griffiths) who was working to re-establish traditional Kamilaroi ceremonial culture with young Kamilaroi men. Greg Griffiths had seen some of the information from the Aboriginal Astronomy Project, which is a cooperative project among academics to find and record the sky knowledge of Australian Aboriginal and Torres Strait Islanders (www.emudreaming.com). Greg contacted Norris and indicated that he had knowledge of Kamilaroi sky culture that he would like to share, and became the first participant in the ethnographic phase of this project.

Prior to the start of the project, the project plan was approved by the Macquarie University Human Research Ethics Committee. As the ethnographic process included interviewing not just the initial participant, but other interested people of Kamilaroi or neighbouring language group ancestry, consultation with the relevant communities was required.

After extensive consultation, interviews were conducted with 8 people. The ethnographic process was similar to that of Mathews and other early ethnographers, being the recording of stories and information from the participants. No specific interview process was designed, and specifically, no list of questions was used, to avoid prompting answers. The participants were simply asked to tell their knowledge of the sky as they understood it linked to their culture, and if appropriate, explain the stories at different levels of cultural knowledge.

Individual intellectual property was protected by the use of an Inform and Consent document between Macquarie University and the individual participant, and by the checking of information received orally through a Story Collection Review Form, which was sent to the participant for checking the information transcribed from the interview.

### 2.3 Data Analysis

The Literature Review resulted in a database which was analysed for references meeting the requirements of (a) relevance to the geographical area of the study, (b) reliability of the source, and (c) whether the story was about an astronomical object. This database was then sorted into sky objects and phenomena, with an entry for such items in the literature which were clearly identifiable. The database included the categories of Star lore, Religion, Art, and Vocabulary.

The ethnographic data consisted of Story Collection Review Forms approved by the participants.  The stories which made up the ethnographic data were sorted into the same categories as the Literature Review, with the sources identified by code, and comments where the story was similar to one in the Literature Review.  This formed the Collected Stories database.

Both databases were analysed and the frequency of objects compared. Objects with a sufficient number of references were put into a combined database of literature and collected stories.  These combined stories were then discussed with the project Reference Group, which consisted of Greg Griffiths (Kamilaroi), and Michael Anderson (Euahlayi/Kamilaroi).  The stories were reviewed object-by-object, and both reviewers were invited to add comments to the stories, either literature or collected.  The results of the review were added to the combined database as guidance for the eventual analysis.

We analysed the combined database and developed a consistent story or stories for each object, and where necessary, further broke down the story by cultural group.  If sufficient information was available, each story was analysed with regards to whether it was first mentioned in early literature, and if it was not, whether it appeared that the story had significant influence from post-colonial contact.  If it was shown that a story had a component of ethnoscientific knowledge, such as understanding of the movement of the objects in the sky, then this was noted and used to answer the Hypothesis 2 on ethnoscientific knowledge.

## 2.4 Giving back

While the academic purpose of this project was to find, record, and publish information about the sky knowledge of the Kamilaroi and neighbouring language groups, a secondary, but very important part of the project is the "giving back" of knowledge to the Kamilaroi and neighbouring communities.  The purpose of this is to allow the communities access to information that may have been lost locally, promote community pride in their culture, and most importantly, provide educational material for young Aboriginal people in the area of study (and beyond).  This phase of the project will take place after the majority of the data is collected, analysed, and published, so that the knowledge has passed through sufficient review steps to ensure that it is accurate and relevant to the community.

# 3. Results

## 3.1 Results of the literature study

The literature study was based on numerous database searches for books, articles, and papers, commencing with the Mura® database at the Australian Institute of Aboriginal and Torres Strait Islander Studies (AIATSIS) Physical reading of 110 potential sources (most sources at AIATSIS can only be read in their library) lead to the researchers developing 55 references that could be used in the study. Some sources could not be used due to cultural sensitivity issues. Further study was carried out through database searches of libraries, including the National Library of Australia, State Library of NSW, and the Macquarie University Library, followed by a general search of scholarly databases, including Google Scholar and JSTOR. It is estimated that approximately 5000 references to the Kamilaroi were at least searched on-line for information on sky culture. Based on the criteria established in the previous methodology section, a total of 125 items were included in the literature database, the oldest being from 1857, and the newest from 2013.

## 3.2 Results of the ethnographic story collection

The ethnographic collection of stories commenced in November 2012, and was sufficiently complete to start the review process in June 2013. A total of eight participants were interviewed, in some cases up to three times. Four participants identified themselves as Kamilaroi, one as Euahlayi, one as Murrawarri, and two as Ngemba. Most participants had heritages from one or more of the other language groups, which is an indication of the "mixing" of cultures and stories in the area of the study. In some cases, stories provided by participants were prefaced by information that the story came from another person, either deceased, or not able to directly participate but willing to provide a story. Participants came from a wide range of places within the area of the study, but many were either born, or had strong links, to the northwest of the study area, in particular, Walgett, Brewarina, Weilmorangle, and Lightning Ridge. All but one participant agreed to being acknowledged in the published literature, but are not identified in respect to specific stories, other than by a "P" followed by a number. After the participants reviewed the individual Story Collection Forms, a total of 105 items were included in the collected stories database, using the criteria established in the Methodology.

## 3.3 Reference Group reviews and the hypotheses

The Reference Group reviewed both the literature and collected stories databases with Fuller, to reach a consensus on the meanings of database items, original sources, and possible cultural connections. Each story was then comprehensively analysed, and rewritten before a final review by the Reference Group prior to publication. Those reviewed stories connect with the 13 astronomical objects which are analysed in the Results, and will be the basis for testing Hypotheses 1 and 2 described in the Methodology.

## 3.4 Selection of combined database

The two databases were analysed as shown in Table 1:

| Object/Starlore | Literature Database | Collected Stories DB |
|---|---|---|
| Aldebaran | | 1 |
| Aurora | | 1 |
| Canopus | 2 | |
| Cluster | | 1 |
| Coal Sack | 3 | 1 |
| Comets | 2 | |
| Corvus | 1 | |
| Dark Patches | | 2 |
| Eclipses | 2 | |
| Emu | 1 | 9 |
| Jupiter | 2 | 2 |
| Kangaroo | 2 | 3 |
| Magellanic Clouds | 7 | 4 |
| Meteors | 5 | 5 |
| Milky Way | 9 | 18 |
| Moon | 9 | 7 |
| Orion | 2 | 5 |
| Pleiades | 18 | 5 |
| Pointers | | 2 |
| Saturn | 1 | |
| Scorpius | 2 | 5 |
| Southern Cross | 10 | 9 |
| Stars | 19 | 2 |
| Sun | 3 | 3 |
| Ursa Major | 1 | |
| Venus | 8 | 4 |
| Vocabulary items | 14 | |
| Culture items | 4 | 16 |

*Tab. 1 frequency of objects in literature and collected stories databases*

The 13 astronomical objects highlighted in grey were chosen for inclusion in the combined stories database. An object was selected for inclusion if it had at least four references in the combined database, sufficient to allow an analysis of the stories of that object. Only a few of the Western constellations appear in the literature and collected stories of this study, and these are the very bright and obvious patterns of stars that make up the Southern Cross, Scorpius and Orion. Orion is a summer constellation, and Scorpius a winter one in Australia. Vocabulary and purely cultural stories (with no reference to an astronomical object) were not included.

## 3.5 Analysis of the stories

The stories collected in this study, like most Aboriginal cultural stories, do not merely entertain and perhaps explain some physical object. Stories are extremely important to cultures with oral transmission of knowledge, and may be the main means of transmitting the Law. Aboriginal customary Law governs all aspects of Aboriginal life, establishing a
person's rights and responsibilities to others, as well as to the land and natural resources (Law Reform Commission of WA, 2006: 64). The use of cultural stories to transmit Law means that most stories have different levels of meaning beyond that of entertainment. Sveiby and Skuthorpe (2006: 45-51) explained four levels of meaning; the first being for children, to explain natural features and animal behaviour, the second concerning the relationships between people within the community, the third the relationship between the community and the larger environment (country), and the fourth the practice of ceremonies related to the story. A participant in this study (P7) said that some stories could have up to "30 levels" of meaning. In this study we have attempted to collect information on stories including the first three levels where available, but have avoided all but general references to ceremonial levels.

The stories of the 13 objects have been analysed to determine whether there were dominant cultural stories about that astronomical object, and based on that analysis, those stories were recorded as close to the original form as could be determined from the sources, both literature and ethnographic. If there was no dominant story, then the cultural significance of the object was summarised and recorded in place of a story. We also recorded the language group from which the story or cultural significance originated, and whether it showed evidence of an ethnoscientific approach

In the following sets of stories, we have grouped related or similar objects. The spelling of proper names and objects follow normal practice in the literature. In the case of collected stories, if the participant was Kamilaroi/Euahlayi, the *Gamilaraay/Yuwaalaraay/ Yuwaalaryaay Dictionary* (Ash et al, 2003) was used with the notations "G/Y", "G", or "Y". If the story was from another language group, the spelling given by the participant was used, if possible.

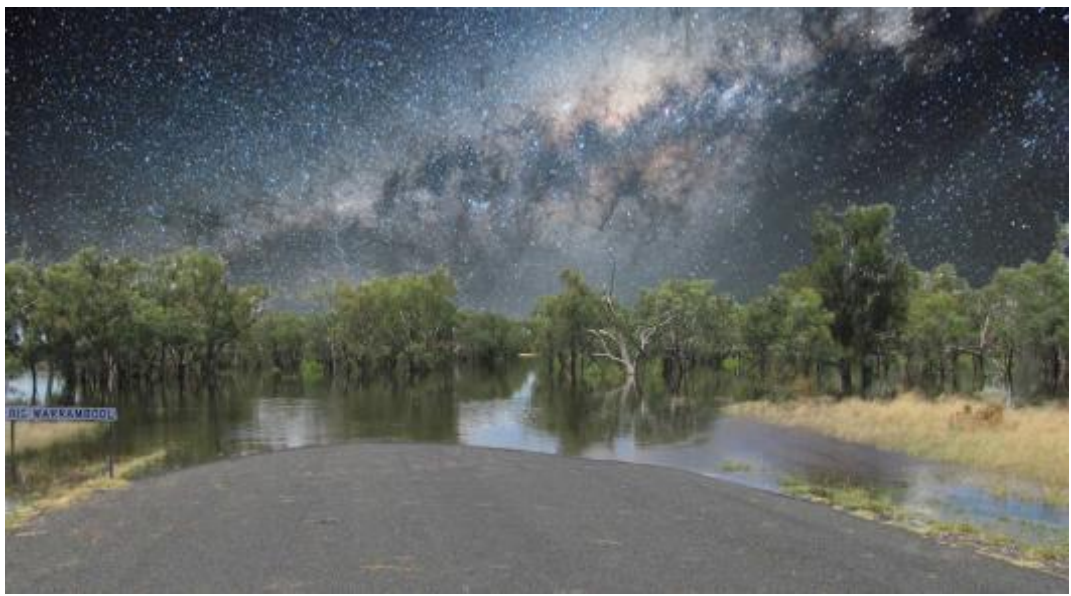

*Fig. 2 Warrumbul in the sky/Warrambool on the ground (based on a sky image by Alex Cherney)*

### 3.5.1 The Milky Way

In Australia from March to October, the Milky Way is a spectacular sight from a dark location, with its myriads of stars and dust lanes. Before cities brought light pollution, Aboriginal people had an unimpeded view of the Milky Way every clear, moonless night, and this study found that it had a significant place in their knowledge of the sky. From the earliest literature (Ridley, 1873: 273-4), it was identified as *Warambul* (G/Y), or *Warrambool*, and Ridley later (1878: 286) translated this as "stream". Parker and Lang (1905: 71-2) further described the Milky Way as "*Warrambool*, or water overflow, the stars are fires, the haze is smoke from them, which spirits of the dead have lit." This last description is disputed by P2 and P4, in that the dead are not along the *Warrambul* but have gone to *Bulimah* (G/Y), which is "behind" the Milky Way (P2) (this is the "sky camp" which is loosely defined by participants as "heaven"). The whole Milky Way is *Warrambul*, the big river in the sky, and has no water, which is caught on Earth (P2).

Surprisingly, for such an important part of the night sky, the Milky Way has only one specific story attached to it by any of the language groups in this study. However, a large part of the stories and culturally significant knowledge of the sky are referenced to the Milky Way/*Warrambul*. P2 says that "the Milky Way represents where things are – campsites, tribes, ancestral places, in other words, a sky atlas, and a big library that defines things." In particular, the Emu, the Kangaroo, and the Crocodile are located there, and their stories follow.

The one story of the Milky Way itself, from P2, is about the sons of the culture hero, *Baayami* (G/Y), who, after his sons disobeyed him, turned them into large rocks which can be seen today on either side of the stone fish traps on the Barwon River at Brewarrina. The two bright patches either side of the Milky Way in Sagittarius (the Galactic bulge) are the spirits of the sons, but the bodies are the rocks.

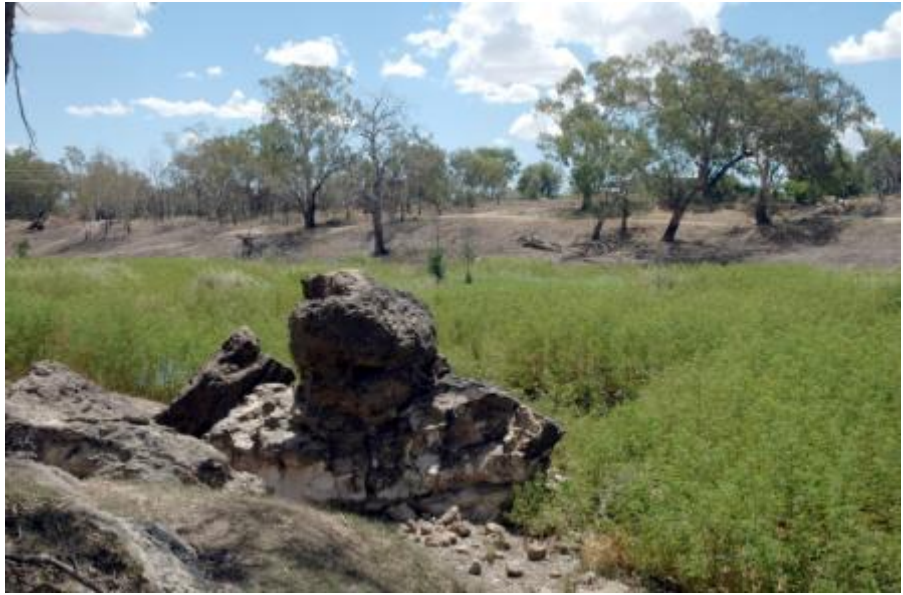

*Fig. 3 one of Baayami's sons at the Brewarrina fish traps (Wikipedia Commons)*

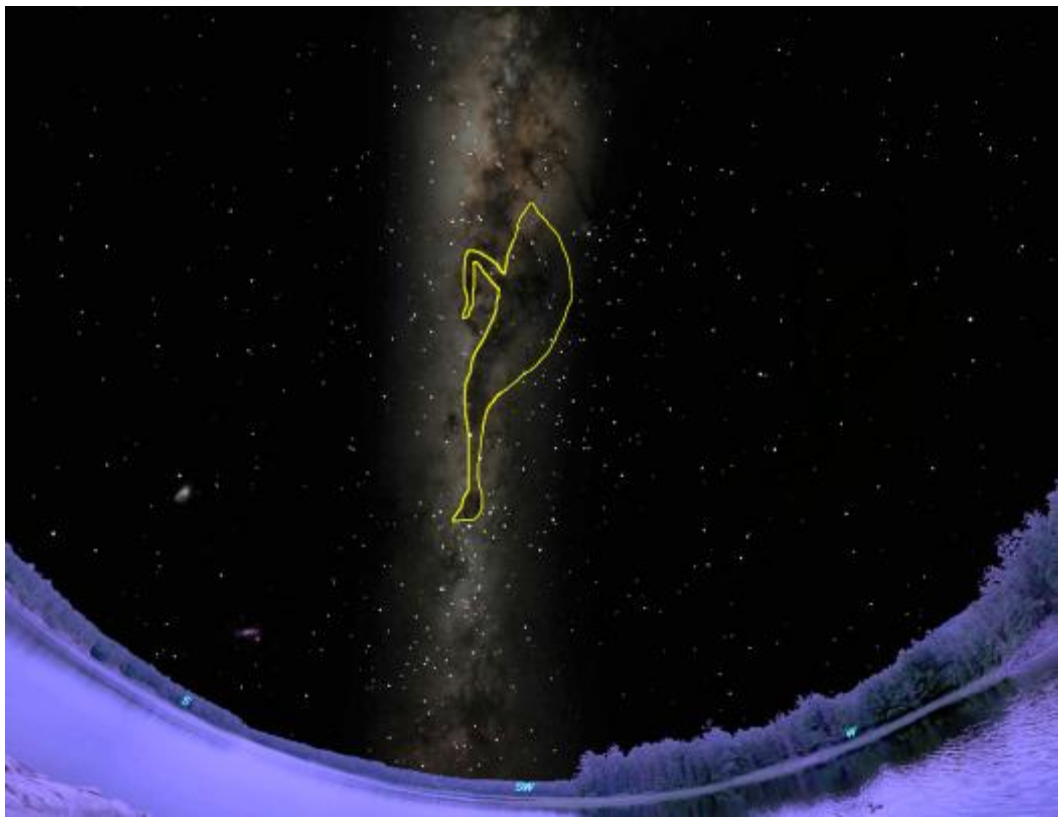

*Fig. 4 the Emu in the Sky early evening in August (image courtesy of Starry Night Education)*

## 3.5.2 The Emu in the Sky

To many Aboriginal language groups across Australia, the Coalsack (which is a dark nebula next to the Southern Cross) represents the head of the Emu in the Sky, an Aboriginal "constellation" which stretches from the head to the body and legs, through the dark lanes in the Milky Way, to the constellation Scorpius (Norris and Norris, 2009: 5-6). The word "constellation" is in quotes, because for many Aboriginal peoples, the dark patches in the sky, particularly the dark dust lanes in the Milky Way, are more important than the patterns of bright stars that form Western constellations. The first mention of the Emu in literature was in Ridley (1873: 273-4), where an Aboriginal person named King Rory informed Ridley of the names of a number of astronomical objects, including *gao-ergi* "emu in the dark space under the tree" (we have taken this to mean that the "tree" is the Southern Cross, and the "dark space" is the Coalsack). Parker and Lang (1905: 73) later told the story of *Gowargay*, the featherless emu, who is the "devil" of waterholes, but goes to his sky camp at night in the Coalsack. King Rory's *gao-ergi* is the same as *Gawarrgay*, which is the current Kamilaroi/Euahlayi spelling and pronunciation. A current source (P2) confirmed that *Gawarrgay* is the ceremonial name for the Emu in the Sky, and that *dhinawan* (G/Y) is the name for the "real" emu bird on the Earth. There are two aspects of the Emu in the Sky in the culture being studied in this paper, the first being stories connected to the Emu, and the second the significance of the Emu in ceremony and keeping of seasons.

A woman's story (P6) tells children to:

> Look in the space between the stars, what do you see? What you're looking at is an emu running, and when it runs, it sticks its neck out. When the old people saw this shape in the sky they knew that the emus were laying their eggs, so the old people would go out into the bush to find the eggs for food. The Emu in the Sky usually appears in April and May (P6; Tribe, 2011: 15).

This story corresponds with the rock engraving of the emu at Kuringai Chase National Park, which aligns with the Emu in the Sky at this time of year (Norris and Hamacher 2009: 13-14). Another level of meaning might be the relationship with country and resources (the emu eggs), which is usually the business of women (P1). P1 said that:

> When the eggs are visible in the body of the Emu in the Sky, that's the time (April) for hunting the eggs. When the eggs push down to the legs (end of June), the chicks are forming, so stop hunting eggs.

The Emu in the Sky is significant in ceremony, although much of this falls into the level of meaning that is not explored in this study. *Gawarrgay* is a featherless emu, and the head in the Coalsack represents a big waterhole on Earth. This emu is believed to inhabit waterholes on Earth and looks after everything that lives there (P2), which may relate to Parker's story of *Gowargay*. P2 also said that spirits like *Gawarrgay* travel from waterhole to waterhole underground and these waterholes are mirrored by dark spots in the sky. We will expand on the subject of places on the ground being mirrored in the sky in the Discussion. Another story (P2) is that "if you can see the head of the Emu in the Sky, the waterholes in country are full, and if it is not visible, they are empty", which may relate to seasons.

> When they are empty the water goes back to the belly of the Emu. The appearance of the Emu in the Sky has connection to movement in country, as well. When the head appears in February, people start moving from their summer camp. When the legs appear in April, people go home to their winter camp. When the neck and legs disappear around August-September, the belly is still there, and represents the egg, which is now developing into a chick. (P2)

Another connection to ceremony is the relationship between the Emu in the Sky and the male initiation ceremony, known as the *Bora* (G/Y) in the area of this study. Another Kamilaroi word for emu is *ngurran.gali* (Euahlayi: *dthnarwon.gulli*), meaning "an emu sitting" or "emu in the water". P4 relates that this means an emu sitting in water, connecting to the Emu in the Sky, whose connection on Earth is the waterhole. Fuller et al (2013) have examined the relationship between the alignment of *Bora* grounds in southeast Australia and the sky, and have found a strong correlation with the southern quadrant. In one of the common times for the *Bora* ceremony, August, the Milky Way/Emu in the Sky is a vertical feature pointing to the south-southwest in the evening (Fig. 4). P4 further confirmed that there was some relationship between the *Bora* and the emu, as male emus care for emu chicks, and elder men take young men through the knowledge ceremony that is the *Bora*.

### 3.5.3 The Kangaroo/Crocodiles in the Sky

As one of the most important resources in pre-invasion Australia, it is unsurprising to find the kangaroo placed in the sky. The crocodile, however, is an estuarine animal found today only across the north of Australia, and is not known historically in the area of this study. We look at both as a part of culture and stories related to the Milky Way.

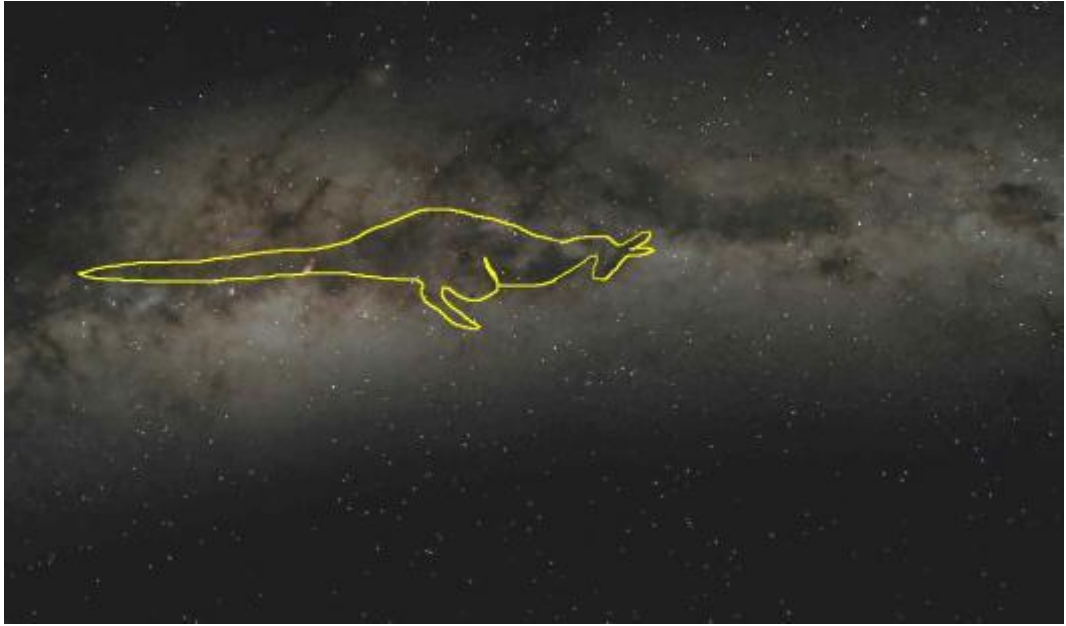

*Fig. 5 Kangaroo in the Sky (under the Emu) (image courtesy of Starry Night Education)*

The earliest references to a kangaroo in the sky are from Ridley (1873: 274-5), who first said the four stars of the constellation Corona Australis were called *Bundar*, the kangaroo. Later (Ridley, 1878: 286) said that *Bundar* was the constellation Corvus. Other than Parker and Lang (1905: 73) repeating the last statement there has been no further mention of a kangaroo in the literature studied. *Bandaar* is Kamilaroi/Euahlayi for "grey kangaroo". The lead author was told by an amateur astronomer that he had been shown the kangaroo in the Milky Way while working in the Kimberley region of Western Australia, and pointed out a shape in the dust clouds of the Milky Way below the Emu in the Sky that could be seen to be a kangaroo. There was no story or cultural significance passed on in this knowledge, but it shows the possible universal nature of this object.

The only mention of a crocodile in the sky is in Parker and Lang (1905: 95), where she describes in the Milky Way a "dark shadow of a crocodile", which seems strange, as she was located in an area (Narran Lakes) where the last known crocodile was *Pallimnarchus*, a crocodile from the Pleistocene, extinct over 40,000 years BP (Gillespie and David, 2001: 42). Parker names it *Kurreah*, and in the current Kamilaroi/Euahlayi language there is a word, *garriya*, "crocodile". Radcliffe-Brown (1930: 342-4), in an investigation of the rainbow serpent in southeast Australia, speculated that *kuria* was actually the rainbow serpent.

In spite of the limited literature recording of the Kangaroo and the Crocodiles in the Sky, there were a number of participants with knowledge of these objects. The Kangaroo was known by all the language groups in this study, with the exception of the Murrawarri, where there was no information available. The Kangaroo was described by participants (P2, P4, and P7) as being beneath the Emu in the Sky, towards the Emu's *bubudhala* "tail" (G/Y), and includes feet, backbone, head, and tail, with the body parallel to the Milky Way, and the head facing away from the Emu. Once it was pointed out on a dark night, it was clear to see (Fig. 5).

> We were told only one story about the Kangaroo, although there may be more stories at a ceremonial level. The public story is about why the kangaroo got his tail. In the old days, the kangaroo (who was a "person") was seen to be disrespecting and mocking the ceremonies, so *Baayami* and the old fellas said that the kangaroo totem was banned from ceremony. Even though they were banned, they joined in anyway, so *Baayami* decided to punish the kangaroo. He said that even though men lost their tails, the kangaroo will keep its tail, and never walk again (P2, P4).

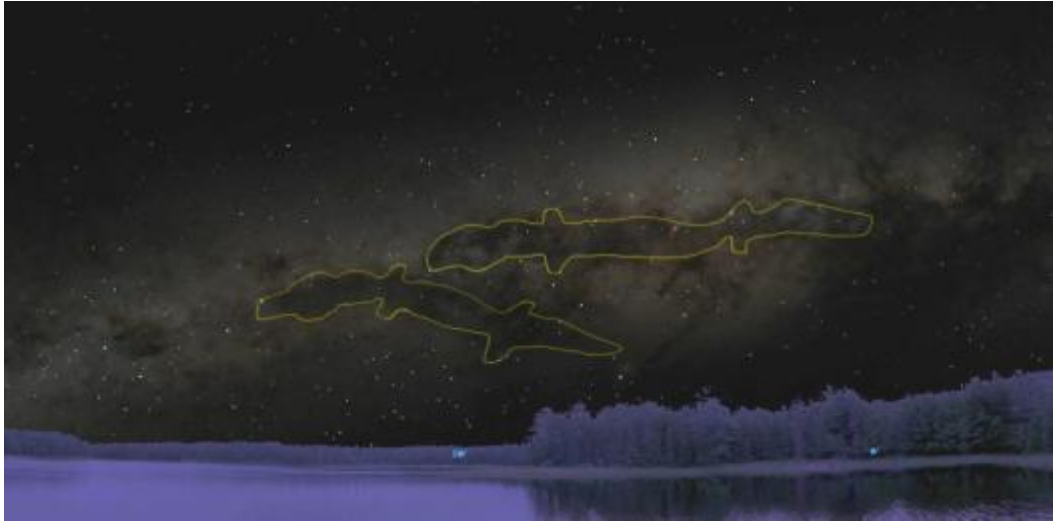

*Fig. 6 Crocodiles in the Sky (image courtesy of Starry Night Education)*

The Crocodile, other than the mention by Parker and Lang (1905), and Radcliffe-Brown's (1930) speculation, is similarly mysterious regarding its place in knowledge of the sky. It doesn't become visible in the Milky Way until late summer, when the Kangaroo, then Emu disappear, and the Crocodile forms in their place, with the belly of the Emu becoming one of the Crocodile's heads (Fig. 6) (P2). The Euahlayi see the Crocodiles as lying in the "water" of the Milky Way, and use it as a signal to start travelling to ceremony in September/October (P2). Like the Kangaroo there is no story specifically connected, but there is a story from Sveiby & Skuthorpe (2006: 61-3) about *Baayami* when he was on Earth, with his two wives, *Birrangulu* and *Ganhanbili* (G/Y). The story is about the forming of Narran Lake. *Baayami* instructed his wives where to find fresh water at Coorigal Springs while he was hunting, but told them not to swim there. They misunderstood, had a swim, and were swallowed by two *garriyas* (crocodiles), which fled through to the Narran River. *Baayami* caught the crocodiles in there, killed them, and freed his wives, who were revived on an anthill. The crocodiles, in dying, thrashed about, and created the Narran and Coocoran Lakes (P2 corrected the location from Sveiby & Skuthorpe). P2 also said that the crocodiles' spirits were punished, were not allowed to go to Bulimah, and must stay on Earth, becoming the protectors of women's sacred grounds. This also was a story of death and resurrection (of *Baayami's* wives), and from this came the ceremonies of initiation for men and women, representing the passing from childhood to adulthood. This is the only story the researchers found from this area which mentioned crocodiles, and given the levels of meaning in the story, it may very well relate to the *kuria* in the Milky Way.

### 3.5.4 The Magellanic Clouds

The Large and Small Magellanic Clouds are small galaxies close to our Milky Way Galaxy. On a clear dark night in southern Australia, they resemble cotton balls, but are made up of millions of stars.

Ridley (1873: 273-4) was the first to comment on the Magellanic Clouds, and was told by King Rory that they were two *buralga* (brolgas – native birds) (G/Y). Parker (1898: 21-7) says they were a mother and daughter who were chased by *Daens* (*Dthane*, the same word, means a Euahlayi man), captured by *Wilbaarr* "the whirlwind spirit" (G/Y), turned into brolgas, and when they died, went into the sky and became the Clouds. Howitt (1904: 439), however, says "that the Kamilaroi believe that the spirit of a man when he dies goes to a dark patch in the Magellanic Clouds, which they call *Maianba*, meaning endless water or river".

In the collected stories, even with many variations, there seemed to be a common theme, which was closest to the account by Howitt (1904: 439); that the Magellanic Clouds were a place where people went after death. Women's stories clearly said (P1, and P3 - a Murrawarri source) the Clouds were openings to heaven.

> P2 and P4 said it was more complex, as the Large Magellanic Cloud is home to *Baayami's* third wife, who sings to women who are going to have babies. The Small Magellanic Cloud is an old man camping, and in all Aboriginal cemeteries on Earth, anyone buried there who has not been initiated, is guided by this old man to *Baayami's* wife in the LMC. This is because uninitiated people can't go to *Bulimah*, so *Baayami's* wife sends them back to Earth as new babies. In all cemeteries in this region is a wilga tree, which represents the old man in the SMC.

## 3.5.5 The Southern Cross and the Pointers

King Rory told Ridley (1873: 273-4; and 1878: 286) that the Southern Cross was *zuu* or *nguu* (tea-tree) (G/Y). This was interpreted by Fredrick (2008: 70) as an "emu sitting under a tea-tree", the emu being *gao-ergi* in the Coalsack. Parker and Lang (1905: 73) expand on this to describe it as the first *Minggah*, or spirit tree, which was the medium for the movement to the sky of the first man on Earth to die. In a later story (Parker, 1914: 8) she further details this as the first two men and a woman, and the death of one of the men, who was lifted to the Milky Way in a giant *yarran* (river red gum) (G/Y). The four stars of the Cross are the eyes of the man who died, and of *Yowee*, the Spirit of Death. The Pointers (stars Alpha and Beta Centauri) are two cockatoos. This story has been repeated by Wolkowsky (1968: 102-4) and Robinson (1968: 77), as well as other collectors of stories. Sveiby & Skuthorpe (2006: 97-8) have repeated a detailed version as follows: "in the Creation time, two men and a woman came from the red country and had been shown which plants they could eat to stay alive. They did this for a long time, but then a big drought came, and they were hungry. One man killed a wallaby, even though he was told by the other man that he should not, as he did not know the law of that totem (the wallaby). He and the woman ate some of it. The other man refused, and walked across the sandhills and the river until he came to a big white gum tree. He lay down and died there. A spirit saw that he did not break the law, and put him in the hollow of the tree, and then lifted the tree into the sky, followed by two white cockatoos screeching, because their roosting place was in the tree. The tree was placed in the southern sky, where it faded and only the eyes of the spirit and the man could be seen, forming the Southern Cross. The two cockatoos still fly after the Southern Cross, and are the Pointers. This is the story of the first death of man. The she-oak trees sigh in the wind, and the gum trees cry tears of blood (gum) to mourn for the first death of Aboriginal people. In the system of levels of meaning in stories, the first level explains how the Southern Cross was created, why the gum tree bleeds a resin, and why the she-oak sighs in the wind. The second level describes the first death and the law about burial (in a tree). The third level explains the law about not killing a totem animal, and a fourth level explains how the spirits take the people to the sky."

The Murrawarri have a very different story, recorded by Mathews & White (1994: 6-7), called Baiame and the Sacred Fires, in which a neighbouring tribe tried to steal the sacred fires. The fires, some Murrawarri people, and the neighbours, were lifted into the sky, and can be seen as the Southern Cross. The Pointers are two guards, *Giduba:mbi* and *Dhadeba:mbi*. This story was not confirmed by any Murrawarri participants in this study, but P2 said that it may be connected to *Baayami's* fire in the *Warrambul*.

All of the Kamilaroi/Euahlayi participants who commented on the Southern Cross presented a story similar to that from Sveiby & Skuthorpe, with minor variations, such as the totem animal eaten. P1 and P2 have stated that the differences in the stories are simply because they are from different cultural groups, but the tree lifted into the Southern Cross is always a river tree. For the Kamilaroi, this would be a *yarran*, for the Euahlayi a *gulabaa* (coolabah) (G/Y), and for the Ngemba, a *nguu* (tea-tree). A story (McKay 2001: 38) from the Walgett area, probably from a Kamilaroi person, was similar, but added some further meaning:

> You will know what it is to suffer (referring to the first death), and when you stand under the she-oak tree you will hear the spirits crying in pain, and if you strike a river red gum, it will shed tears of blood. Live in harmony with the Earth and it will give you what you need to survive."

P2 also added a further level of complexity to the meaning of the Southern Cross to reinforce his statement that the sky is a roadmap, and tells where everything is:

> Extending stories to the ceremonial, the sky can be turned into a totally different map. The sky can take one from the northeast in the sea (Pacific), down the coastline to Byron Bay, over Lake Eyre, Warburton, Pilbara, Kimberleys, back down through the Tanami Desert, to Uluru, and back to Lake Eyre. It's a complete roadmap. Lake Eyre is where *Baayami* finished the creating, when he and his wives were travelling. On the way, one of his wives, *Birrangulu*, died, and her spirit stayed at Uluru, to sleep. Her physical body went to Lake Eyre with *Baayami* and his other wife, *Ganhanbili*. At Lake Eyre they went south, and the Southern Cross is a coolabah tree. In Angledool (NSW) is the largest coolabah tree; that tree represents the Southern Cross. That is a dead coolabah tree. That's where *Baayami* and his other wife camped on the way to *Bulimah*. Down here, we're looking up at the black hole on the side of the Southern Cross (Coalsack); that's the hollow/bottom of the coolabah tree. Go up the hollow tree, and when you come out the other side you're home (*Bulimah*), all the people are there. The big star to the east of the Southern Cross (Alpha Muscae) is a fireplace, and all mothers on Earth who have children go to that place when they die. They make camp, and stay there until their last child comes, when they leave for *Bulimah*. Their female children who have children will stay until their last children arrive, and then they will join their mother in *Bulimah*.

P3 also mentioned Angledool in his version, calling it *Yeranbah*, and said that the waterhole there, *Bilambulaa*, was where the crocodiles who took *Baayami's* wives first came up from under the ground.

Given the use by non-Aboriginal Australians of the Southern Cross and the Pointers for finding south, it was surprising that there was only one mention of this by P1, who said that the Pointers give you directions where you go, and how to go home.

### 3.5.6 The Pleiades star cluster

The Pleiades are an open star cluster which can be seen in the northern sky during summer. Krupp (1994: 86) says they are a very special object in the sky, which is recognised by nearly every culture as something special. In some older cultures, such as the ancient Greeks, the Berbers of North Africa, and the Lakota Sioux of North America, the Pleiades were seven women, young women, or "daughters".

There are several Kamilaroi stories about the Pleiades. Ridley (1875: 24-6), said the Pleiades were *Miai-Miai*. *Miyay* means "girl", *miyay miyay* (several girls) (G/Y). Greenway (1878: 243) said the Pleiades or *Miyay Miyay* lived on Earth, and were exceptional beauties. He also said that Orion, or *beraiberai* (*birray* means "boy", *birray birray* "several boys"); being uninitiated boys or young men, pursued them, and the *Miyay Miyay* prayed for deliverance. *Bhaiami* and *Turramulan* granted their request, and they were lifted into the sky. One is not as beautiful as the rest, and hides behind (most people only see six stars [Kyselka 1993: 174]). *Beraiberai*, as leader, also went into the sky, and is Orion, with his *burran* (boomerang) and *ghutur* (belt). Greenway (1901, p. 190) later has a Kamilaroi version of this story, with *Werrinah* (G/Y *Wurruna*), a clever man, stealing some of the sisters. Mathews (1904: 280, 283) added a Ngemba story that when the Pleiades rise around 3 or 4 AM, old men take glowing coals from the fire, and cast them towards the Pleiades to prevent spirit women from making it too cold. He also said that the Pleiades were a group of young women searching for yams and a whirlwind put them in the sky. Parker also collected Pleiades stories; the first (Parker, 1898: 22-5) was about *Wurrunnah* (a clever man) and the two of the seven sisters he tried to keep. In the second (Parker and Lang, 1905: 72), Orion's Belt is the *Berai-Berai*, boys who love the *Meamai* (Pleiades) but were rejected. They died of love, and the spirits put them in Orion where they hunt by day and dance to a cooroboree at night (music from the Pleiades). This last story is similar to the story that Stanbridge recorded from the Boorong people of northwest Victoria (in Smythe, 1898, Vol. 1: 430-4) in which Orion is *Kuckan bulla*, young men dancing, and the Pleiades are *Larnan kurrk*, a group of young women playing to *Kuckan bulla*. This story may be common throughout southeast Australia.

Later writers, including Mountford (1976), D'Arcy (1997), and Clarke (2007), have similar stories about the Pleiades. A different version in the literature was from Johnson (1998: 116), who said the *Mayi-mayi* were seven sisters with long hair and bodies of icicles. A large family of young men, the *Berai-berai* (Orion) followed them. Thunder in the winter is the Pleiades bathing and playing (ibid: 116). Haynes (in Selin, 2000: 78) tells a Kamilaroi story of *Meamei* or *Mayi-mayi*, seven sisters with long hair and bodies of ice. "Before leaving Earth they travelled into the mountains causing springs to feed rivers so there would be water forever. A young hunter, *Karambal*, fell in love with one sister and carried her off. Other sisters sent cold, wintry weather to force him to release her, but later relented and made their way into the sky in search of the summer sun to melt snow and ice. Thus the Pleiades appear in the summer each year, bringing warm weather. Afterwards they travel west and winter returns as a reminder that it is wrong to carry off women who belong to a totem forbidden them. *Karambal* ascended with them and still pursues them as the star Aldebaran, which follows closely."

Sveiby & Skuthorpe (2006: 114-7), tell the story of *Wurunna*, who left his community to search for another group with which to live. He came to a camp, and found seven girls, who fed him and allowed him to camp. Their name was *Mirrai Mirrai*, and they were visiting that country. Through various means, he managed to steal two of the girls, and their sisters did not return to find them. *Wurunna* told the two sisters to cut some pine bark to start a fire, and they said they must not, but he persisted. When they drove their axes into the bark, the tree began to rise and grow taller, taking the girls with it. *Wurunna* called for them to come back, but the tree reached the sky, and the girl's sisters reached down and helped them up, where they all are to this day. Sveiby & Skuthorpe explain the levels of meaning for this story; the first described how the star cluster was created and the second and third explained the marriage law and the reasons for it, to prevent men from stealing women by force from other communities. (P2 said this story is not from the Euahlayi area).

Some participants seemed to know the stories of the Pleiades, but couldn't add significantly to the literature. P2 and P4 both said that the Pleiades were young women, and in one case Aldebaran was chasing them, and in the other, Aldebaran was protecting them from the young men in Orion (the *birray birray* are Orion's Belt). In all of the above versions of the story, the "V" shape of stars next to Aldebaran (the horns of Taurus) is the old man's *gunya* "hut" (G/Y). P2 added to the second story that "Rigel was the fire of the *birray birray*, and the Sword of Orion was their fire poker." P2 also said that the old man (Aldebaran) is also a rainmaker, and creates the rings around the Moon (see the Moon section).

P4, a Kamilaroi man, told the following family story of the Seven Sisters:

> A long time ago there were seven sisters who always wanted to go to the big waterhole in the river. The old women said they should not go there, because there might be wiringins (clever men) to grab them and take them to the sky. The seven sisters had special powers, but the wiringins could take them away. The sisters didn't listen and went to the waterhole. Two wandered away to search for mussels. A wiringin grabbed them and jumped back to his home in the sky. When the sun went down the eldest sister looked for the two, and tried and tried, but couldn't find them. Then they remembered the warning about the wiringins. The five sisters became angry and flung themselves into the night sky. They thought the two missing sisters had left them all the little fires in the sky to show them the way to find them. You can always look up into the night sky and you can see the five sisters looking for their two lost sisters among all the little fires. All the little fires up there in the night sky are what made the Milky Way. The eldest sister is the one right out in front showing the way, and the other three are keeping a lookout in the other directions. The youngest sister is protected by the older ones. The mother comes up and shows her bright light every evening and every morning. She is looking for her daughters. She is the brightest star in the sky and she is always hoping her daughters will know that it is their mother and they will all come to her. When it rains it is the mother star crying for her lost daughters. Mum and Dad would tell us this story in the summer time, when we slept out on the grass, rolled up in our blankets. They would ask us to help the mother out by trying to find the two lost sisters in the sky.

The mother of the Pleiades is apparently the Morning and Evening Star, Venus (P4).

A Ngemba participant had a complete story of the Seven Sisters. He said:

> A clever man had six sister wives, but the seventh sister was not yet a woman. He wanted her, but the older sisters said 'no'. While they were asleep he went underground and tried to rape the seventh sister, but missed. All the sisters were running around, and he was chasing, leaving rock pillars everywhere (such as the Breadknife in Warrumbungles NP). He eventually chased them to South Australia, where he caught the seventh sister, who died. The other sisters fled to the sky, and the clever man chased them and became the Moon. An old clever woman chased him, and became the Sun. When both are in the sky, she's teaching him the Law. This is an important story about underage women. When the Moon is small, that's his shame. He gets small after the Sun rouses (slang: works him up) on him. He's paying back by being the keeper of little girls (the Moon gives girl babies to women). This could also be a story about wrong-way marriage.

There seemed to be a connection between songlines and the Pleiades (songlines are stories that have connections to paths on the ground and in the sky). P2 said the Euahlayi have a Black Snake story/songline to the south, where the mother and father of the Pleiades are at the Snowy Mountains.

On a practical matter, P1 said that if you mock the Pleiades, you'll get a cold wind and sleet.

### 3.5.7 Scorpius (the constellation)

With a few exceptions, Scorpius is referenced not for the star pattern, but for some dark spaces around it. King Rory (Ridley, 1873: 273-4) said the dark spaces in Scorpius are *Wurrawiburu* (demons), but also said (ibid: 274-5) that Scorpius is *Mundewur* (notches cut in tree for climbing). This could be *mubirr* (G/Y), which is either initiation scars, or carvings, possibly in trees. Later, Ridley (in Smythe, 1878: 286) said "two stars across the Milky Way near Scorpius are *Gijeri-gu* (small green parrots)". Parker and Lang (1905: 72) said "to get to *Warrambool*, the *Wurrawiberoo*, two dark spots in Scorpius, have to be passed, which are devils to catch spirits of the dead". Mathews (1904: 283) said Alpha Scorpii (Antares) is an eaglehawk "wedge tail eagle", and later (1905: 81) he said Scorpius is an eaglehawk.  P2 has said that the references to an eaglehawk are likely to be Murrawarri in origin.

P2 gave quite a few references to Scorpius in Kamilaroi/Euahlayi culture.  The one that has the most connection to the literature concerns the "dark spots" in Scorpius, which are identified as being in the tail and "claws" of Scorpius.  These are not obvious in Scorpius at night, but in astrophotography images there are two or three dust spots or dark nebulae in the vicinity of the tail and claws (Fig. 7), so it's possible that sharp eyes could distinguish them.

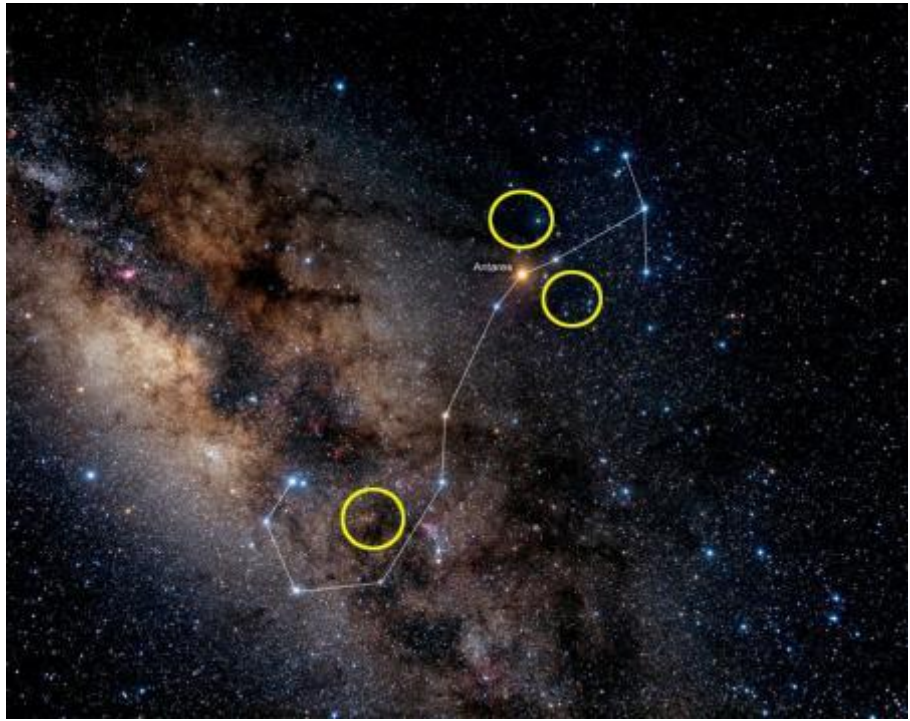

*Fig. 7 dark spots in Scorpius (image: (abc.net.au/science)*

P2 said these dark spots are actually holes, which the spirit of the whirlwind, *Wilbaarr* (G/Y), uses to come to Earth in September, when whirlwinds are common (known as dust devils; the Australian "willy willy" seems to come from the Euahlayi word *Wirrawilburro*).  In the northwest of NSW, September is the windiest month, and whirlwinds are most likely.  P2 said:

> *Wilbaarr* has a reputation for creating madness and stealing souls.  *Baayami* eventually calls him back, but can't stop him coming to Earth, as he can come out through any of the three dark spots.  September is also the time the sacred fire is lit, so young men travel, and *Wilbaarr* tries to catch them.  For this reason, pregnant women, and women with children avoid whirlwinds.

P2 also showed us how the ground in the Narran and Barwon River countries is connected to the three dark spots.  Three depressions in the ground, one each in Euahlayi, Murrawarri, and Kamilaroi country, are called *buulii* (G/Y).  The fact that there are three *buuliis* on the ground reflects the theme that what is up in the sky is also on the ground.  *Buulii*, in Kamilaroi/Euahlayi, also means "whirlwind", so the connection with *Wilbaarr* is clear.  The *buuliis* in this country are located in the same pattern as the *buuliis* in Scorpius, as shown by the Scorpius constellation overlay to the map, which is centred on Walgett.  The *buulii* shown in the inset to Fig. 8, which is approximately 5 kilometres in diameter, was not obvious while driving across it, except that it seemed to be ringed by a circle of trees.  The inset shows a circular shape within the blue circle, which may be related to drainage and moisture.

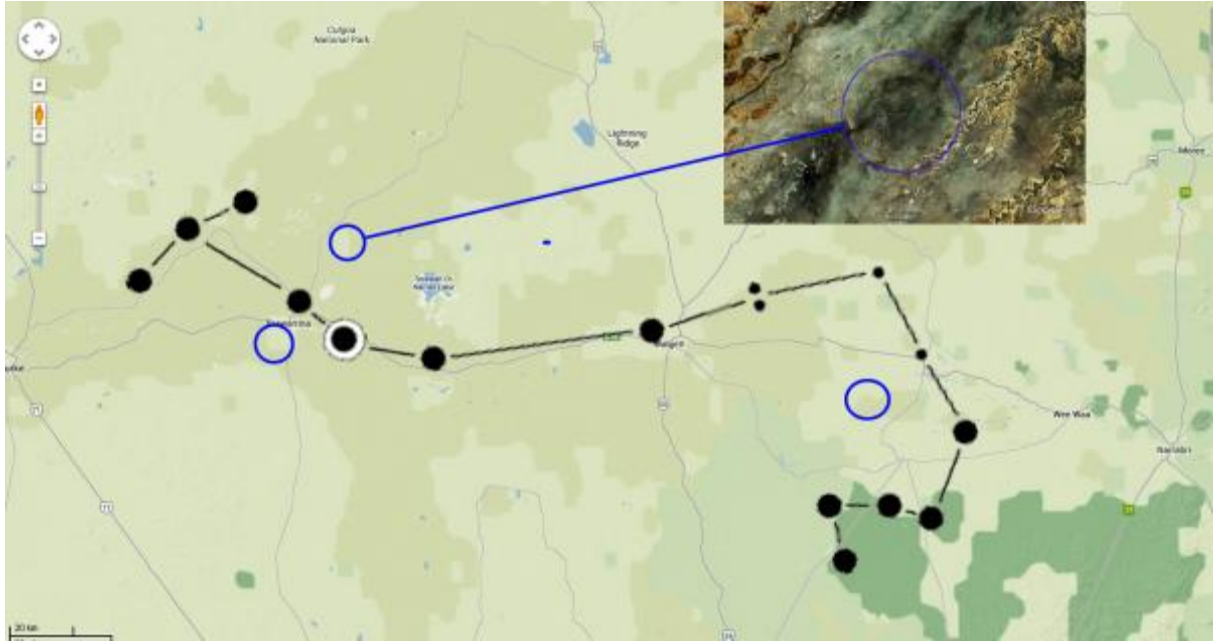

*Fig. 8 the "buuliis" from Scorpius in country (image/map: Google Earth)*

A reference to an eaglehawk in Scorpius was explained by P2, who said "there was an eaglehawk 'landing' if you looked at the shape of Scorpius, which is the head and outstretched legs, and connected to the Southern Cross, which is the wings and tail. This is seen in July and August."

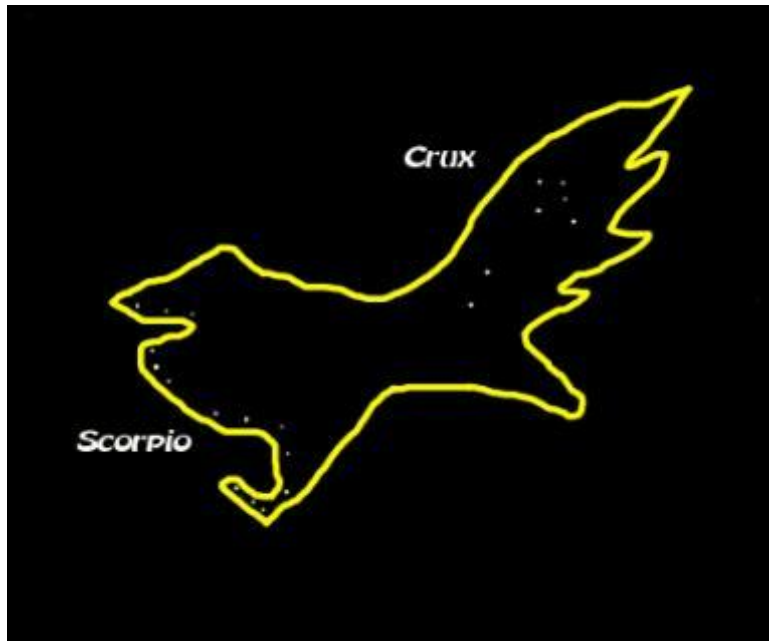

*Fig. 9 Eaglehawk outlined by Scorpius and Crux*

P2 also said that:

> The spirit eaglehawk is *Maliyan* (G/Y) who has a songline that runs East/West, from Alice Springs, where he was killed by *Yipirinya* (a green caterpillar), to Carcool (near Narran Lake). This crosses the songline previously discussed of the Black Snake (and *Waan*, the crow). Where the songlines cross sets up four sections in the sky, which are related to the section or skin system of the Kamilaroi/Euahlayi (and other language groups in southeast Australia). The North/South songline also separates the Dark people from the Light people in the communities, Dark being to the West (sundown), and Light being to the East (sunrise). This is related to the marriage system.

P1 told a women's story about Scorpius:

> When you see the Scorpion, it's the dry season (real scorpions love dry season) – don't pick up sticks then. This is actually true. Old people sent scorpion into the sky because it used to sting people, which is why we don't see many of them.

The wood or forest scorpion is found throughout southeast Australia.

The Ngemba people have a very different story about Scorpius. P7 said:

> Scorpius is the 'broken-neck' brush turkey. The tail of Scorpius is the head (twisted back); the head of Scorpius is the fan tail. The turkey is an important part of Law and ceremony. The brush turkey scratches the ground to make its nesting mound around June, which is when they have the *Bora* ceremony." P7 said that the turkey has been engraved in rock in places, and there is an engraving of the turkey with a clever man pointing to it in the sky, telling a story which is a prophecy.

He also said: "meteors come from Scorpius, and the turkey's neck was broken by a meteor." Investigating this, there are two meteor showers related to Scorpius, the Alpha Scorpiids, peaking in early May with the radiant (apparent source) near Antares, and the Omega Scorpiids, peaking around June 5, with the radiant near the double star, Omega Scorpii. The Omega Scorpiids are reported (Plotner, 2009) to be a bright shower with numerous fireballs.

## 3.5.8 Orion (the constellation)

Orion is a summer constellation that was described as the Hunter by the ancient Greeks, with the four bright stars being his arms and legs, the three stars in a line being his Belt, and the two stars and the Orion Nebula being his Sword.

The early literature, including Ridley (1875: 24-6), Greenway (1878: 243; 190: 168), and Parker and Lang (1905: 72) are unanimous in stating that Orion is the *Berai-Berai* (in various spellings), the young men who loved the Pleiades, and pursued them. Later writers, such as Mountford (1976: 461) and Haynes (in Selin, 2000: 78) also say that Orion was the *Berai-Berai*. The only different story is from Mathews & White (1994: 46) who said that the Murrawarri recognised that Orion wore a belt, carried a shield and stone tomahawk, and called him *Jadi Jadi*, which means either "strong man" or "cyclone".

P2 confirmed that the Kamilaroi/Euahlayi see Orion as the *birray birray*, from whom the old man (Aldebaran) is protecting the Pleiades.

P4, a Kamilaroi, explained that "Orion's belt is *Baayami's buurr*, the hair belt, made by women to symbolise the umbilical to the mother. Orion is showing us how to depict *Baayami* and boys (initiates), but only for ceremonies (Orion is <u>not</u> *Baayami*)."

A women's story from P1 is that:

> The old Saucepan (eight stars in Orion that make up the asterism called the Pot or Saucepan), when it gets full, will turn and tip. The rainy season is usually February in the area of the study, and P1 says it gets full before then (the saucepan asterism appears level in October-November), and tips in February (it is tilted in the sky), as it is full.

P4 said Orion is holding the *birray birray*, and the belt is *Baayami's buurr*.

P7, reporting Ngemba beliefs, <u>does</u> see Orion as *Baayami*. P7 is more detailed in the description, and said that:

> The belt is his *buurr*, Rigel is his head, the long line of stars east of Aldebaran's *gunya* is one arm, and the other arm comes south from the sword, then up as the stars make a boomerang (in the constellation Leptus). The legs come down from the belt. Also, *Baayami* as Orion comes to the ground in June (for the *Bora*), so the *Baayami* ground sculpture, later used in ceremony, is prepared then.

(Orion does go below the western horizon just after it is dark in early June).

P8 had another description of Orion, possibly from the Wailwun (connected to the Ngemba). In this it was said that Orion was the brolga put into the sky by *Baayami* after fighting with the emu (when the Sun was created). P8 said that Orion could be a brolga, emu, or turkey, depending on what country you came from, and birds were put into the sky to show that we shouldn't fight and should help each other.

One of the participants said: "Orion can be lots of things".

### 3.5.9 The Sun

Ridley (1873: 286) reported that the Sun is *yarai*, which is very close to the Kamilaroi *yaraay*. He also reported that "star" is also *yarai*, which if correct, would make his source very knowledgeable (as the Sun is a star) but this is unlikely. Greenway (1878: 236-7) wrote that the Sun is *yarrai*, also Kamilaroi, but that stars were *miri*.

Parker and Lang (1905: 73) collected a Euahlayi story that the Sun, *Yhi* (close to Euahlayi *yaay*), was a wanton woman, and overtook her enemy, the Moon, and tried to kill it, but the clever men prevented it. An alternative story (ibid: 73) was that *Yhi* tried to ensnare *Bahloo*, the Moon, but he wanted none of it, and she chases him across the sky. Fredrick (2008: 102) felt that this story was an explanation of why eclipses occur; that the Sun chases the Moon across the sky after he refuses her advances, and the eclipse is caused by the Sun overtaking the Moon. This story, along with stories from the Wirangu and northwest Arnhem Land peoples, make up the literature examples in Aboriginal culture that appear to be ethnoscientific descriptions of solar eclipses (Hamacher and Norris, 2011: 128).

Reed (1965: 58-62) repeats that the Sun goddess was in love with the Moon, and describes (Reed, 1965: 15-8) the story of *Yhi*, the Sun, and the bringing of life to the world. She also sent the Morning Star to herald her coming each day, and then gave the Morning Star to *Bahloo*, the Moon, for company.

The Morning Star has some significance in stories about the Sun, as discussed in Section 3.4.2, when Parker and Lang (1897: 7-8) reported that the Morning Star was sent to warn people that the Sun was about to rise.

P7 tells a story similar to that of Parker and explains how the Sun was made.

> In this story, there was no Sun, only Moon and stars, and only animals, no people. Either the brush turkey or brolga, depending on the version, after some business with the emu, got angry, and went to the emu nest, and took an egg. She threw it to the East, where it hit a pile of logs, which burst into flames (Parker was reporting a Euahlayi version where the brolga threw the emu egg up into the sky where it hit a pile of brush). The flames lit up the country/world, forming the Sun. After the Sun went across the sky for the first time, that was the night, with the Sun returning each day. A variation says that a spirit in the sky saw how bright and beautiful the world was, and that each day he made a fire after collecting wood every night. When the heap was big enough, he sent the morning star (Venus) to warn those on Earth that the fire would soon be lit. In this Ngemba story, the Creator scolded the turkey/brolga and the emu for fighting, so the turkey/brolga was put in the sky near Orion, and the emu was put with its head in the Coalsack, which is the Emu in the Sky. Other levels of meaning for this story are unclear, beyond one of relationships and not fighting with others, but the Emu in Sky clearly has significance in ceremony and keeping of seasons.

P4 added that the kookaburra (*gugurrgaagaa*) laughed at this event, and reminds people of the story ever since by laughing at the dawn.

P8 repeated the story of how the Sun was made, and P2 added the cultural story that if you let your (sun) shadow fall on someone else's shadow, you give them more life, which is deducted from yours.

### 3.5.10 The Moon

Surprisingly, neither Ridley nor any of the early collectors in the area of the study wrote about the Moon. Parker (1898: 8-10) first wrote about *Bahloo* (*baaluu* Y, *gilay* G) in a story where *Bahloo* tried to convince the first men (*Daens*) to carry his dogs across the river. He showed the *Daens* that it's safe, but when they refused, the dogs became snakes, which bit them, and *Bahloo* said that men will stay on Earth, and die, which is why men kill snakes whenever they see them. Parker also had a story about *Bahloo* and *Wahn*, the crow (ibid: 31-34). *Bahloo*, the Moon, was the maker of girl babies, and *Wahn* sometimes helped. *Buumayamayal* (Y), the fly-catcher lizard, made the boys, sometimes with help from *Bahloo*. *Bahloo* and *Wahn* were clever men, and lived together. *Wahn* wanted, as well as making fresh babies, to bring dead people back, but *Bahloo* refused. *Wahn* got angry, and one day saw a big gum tree with grubs, and proposed that *Bahloo* and he get them. *Wahn* stayed on the ground, and *Bahloo* went up in the tree. While he was up there, *Wahn* breathed on the tree and it grew up into the sky, where *Bahloo* then stayed and travelled across it. He wanted to get back to Earth, but he had rejected *Yhi* (the Sun woman), and she prevented him from going back with the help of spirits. *Bahloo* did find a way to trick the spirits by taking the form of *Gawaargay*, the spirit emu (Emu in the Sky). On Earth he got back into making little girls while *Buumayamayal* made more boys. They sent them to *Wadhaagudjaaylwan* (G/Y), *Baayami's* wife in the Large Magellanic Cloud, and she eventually sent them back to Earth as a spirit to be born. Parker said that the local people (Euahlayi) said that if the Moon was late rising, he'd been making girl babies, and they knew when he was coming by the haze that precedes the Moon, saying "*Bahloo* is coming, there is his dust".

Parker and Lang (1905: 73-4) said that the Euahlayi, when they see a halo around the Moon, say "going to rain; *Bahloo* building a house to keep dry".

Mathews & White (1994: 58-60) tell a Murrawarri story: *Gien*, who was a handsome young man who drowned and was revived, then massacred those who left him to drown. When survivors identified him as a murderer, he escaped into the sky, and still lives there (as the Moon). During lunar eclipses, the colour of the Moon is often red, which is his blood. According to Mathews, the Ngemba have a similar story, and use the same word for the Moon.

P2 has a story about *Bahloo* and *Wahn* which is similar to Parker's:

> When *Wadhaagudjaaylwan* sings to a girl, she sends the child's spirit on the rays of the Moon. She directs the rays to a female bumble tree. A woman brings the intended girl, lays her down, sings to *Wadhaagudjaaylwan*, and rubs the girl's belly (preparing a place for the child – she's not actually pregnant yet according to culture). The spirits travel in the dark (shadows), which is why you don't stand on Moon shadows. Orange/red fungi on logs are spirits who didn't get to their mother, and are sung back to *Wadhaagudjaaylwan*.

This resembles the Ngemba story told by P7 in Section 3.5.6 about the Pleiades, and the clever man who ended up in the sky as the Moon.

Most of the current stories are more in the nature of explaining events with the Moon. P1 said that her father told her:

> Look at the Moon – if there is a ring around it; count the stars in the ring; 2, 3 or 7. Why Dad? If there are 3 stars, there will be rain on the 3$^{rd}$ day.

P2 said "the Moon's rays are important in the spirit world. Children shouldn't stare at the Moon, as the Moon's rays will send them mad." P2 also said:

> The Moon sometimes has a ring around it, or even a double ring. The clever man (Aldebaran), who keeps the *birray birray* from the *miyay miyay* (Pleiades), is also a rain man. He sets up a cooroboree ground, and when he starts the cooroboree, you'll find that the Moon represents the stick that goes in the ground, the rainstick, and around the ring, that's the dust kicked up by the dancing. The dust creates the clouds, and brings rain.

## 3.5.11 Jupiter (the planet)

There are several pieces of evidence that the Kamilaroi had noticed that Jupiter wandered (i.e was a planet) relative to the fixed stars.

Tindale (1983: 367) reported a story from the Euahlayi which Fraser (1888: 8-9) also recorded from the Kamilaroi: "In the grasslands of the eastern riverine corridor west of the Great Dividing Range, peoples of several tribes have stories based on the idea that Jupiter is a young boy wandering about the heavens. He is much disliked by his mother, the Sun, so much so that she sends men to spear him at a time when he is moving low down in the western sky. In general the fear of people is that in dry years the grasses may not set seed, and if the Sun woman succeeds in injuring her son this will be sure to happen. An even greater fear is that if the boy were "killed" all people would become ill, would develop blindness, and many would perish. Even *Kukura*, their Moon man, could go blind. Such ideas appear to reflect their own experiences with drought and with the effects of severe malnutrition caused thereby." P2 commented that this story came from the Lake Cochran area of Euahlayi country.

Earlier, Mathews (1904: 283; 1905: 81) records that "Jupiter was a Kilpungurra man called *Wurnda-wurda-yarroa*, who lived on roasted yams, hence got his reddish colour from the fire." Several participants (P2 and P4) said they thought this was a Murrawarri story, and we speculate that the story may refer to Mars, although Jupiter can appear to be "reddish colour". Johnson (1998: 84) said that this story was from Aboriginal groups on the Darling River, which is more likely Murrawarri, and she also suggested the story was about Mars. P2 said that Jupiter was a "red-eye fella. Kids don't play with fire; red-eye fella will follow you and stay all winter." This was an admonition to kids not to play with the campfire.

The one participant story (P3) was that Jupiter was a wandering spirit. "Nan said 'that big star is watching you. Bad spirit. Do you if you're alone.'" P3 added that Jupiter was a bad person.

Jupiter was recognised as a wandering star by people in the area of the study (Tindale, 1983). Jupiter does move from place to place in the sky over the period of its visibility during the year compared to the background stars, and it takes over 11 years to return to the same position at the same time of year.

### 3.5.12 Venus (the planet)

Ridley alone has four different words for Venus, *Zindigindoer* (Ridley, 1873: 273), *I/Jaje-kindamawa* or *I/Jindikin-dawa* (Ridley, 1875: 24.-6), and *Ngindigindoer* (Ridley, 1878: 286), the first and last translated as "you are laughing". Mackenzie (1874: 250) reported *Yindigindiwa* and *Yalgindowa* as "you laugh" and "I laugh" in the Kamilaroi language; also being the names for Venus. Johnson (1998: 84, 116) also describes Venus as "you are laughing" or the "Laughing Star". Parker and Lang (1905: 71) have a story about Venus being the laughing star, a rude old man, who scintillates (twinkles or laughs). There was a detailed story collected in this study about the rude old man and why he was laughing, but cultural issues prevent us from discussing it here. In any case, the fact that Venus, as both the Morning Star and the Evening Star, is often very close to the horizon, and does scintillate (it twinkles due to the light passing through the thicker atmosphere of Earth near the horizon). Mathews & White (1994: 46) also had a Murrawarri story that Venus is laughing (changing colour), but her earlier relative, Mathews (R.H.) (1904: 283) told a Murrawarri story that Venus was a man, *Mirnkabuli*, who lived in a *gurli*, or hut, and lived on mussels and crayfish. The G/Y verb "to laugh" is *gindama-y* (Ash et al, 2003), which appears similar to parts of the names for Venus reported by Ridley and Mackenzie, described above.

Parker (1898: 31-2) tells a story of *Mullyan*, the eaglehawk, who became *Mullyangah*, the Morning Star. Reed (1965: 79-82; 1999: 73-6) has expanded on that story of *Mullian*, the Eaglehawk, who lived in a giant *yarran* tree near the Barwon River, and hunted people for food. Some young men managed to set his home on fire, and he died, becoming *Mullian-ga*, the Morning Star.

These last stories in the literature were confirmed by some of the participants. P3 told a story from Ted Fields (Walgett): "see Venus in early morning (*Muliyan-ga*) and evening (*Mil-muliyan*). Owl at night, and eagle in day. Waterholes (*Mil-muliyan*) at Morgan's Wells near Glengarry opal fields." P2 cleared up the confusion by explaining that "during the day, *Maliyan's* eyes (the eaglehawk) are the eyes of *Baayami*. During the night, *Maliyan's* eyes are Venus and Mars, which become the eyes of *Baayami* at night. Because one is red, and one is blue and green, two rocks are brought together for ceremonies: one is red (opal) from Quilpie, QLD, and the Euahlayi have the green and blue one (opal). When the stones are put together, they are Venus and Mars on Earth. Another location for Venus and Mars on Earth is a place called Mordale near Narran Lakes. These are waterholes side by side, called *Mil-maliyan* and *Maliyan-ga*."

In the G/Y Dictionary, the Morning Star/Venus is *Maliyan.gaalay*, so the Morning Star identity of Venus is *Maliyan-ga/Maliyan.gaalay*. In the same Dictionary, however, the planet Venus is *Nganundi Gindamalaa* (G) or *Murrudhi Gindamalaa* (Y) "he is laughing at you", so the split personality of the Morning Star/Venus is now clear: the Morning Star is the eaglehawk *Maliyan-ga*, but Venus as a star is the Laughing Star.

The Morning Star has a much deeper meaning in Kamilaroi/Euahlayi culture, and that is reflected in the sacred Morning Star Ceremony. Here we report only those details that are not culturally sensitive. P2 reported that the Evening Star (also Venus), when it appears, is a sign to light the sacred fire, which is relit every evening until the Morning Star is seen, at which time the ceremony takes place (and the sacred fire is doused). This ceremony exists beyond the area of this study, and has been described in Yolngu culture (Norris and Norris, 2009: 18-22).

As in the Yolngu version, a wooden pole is used in the ceremony. However, in this case it is used horizontally, and represents the bridge between the dark and light people (the moieties), and honours the marriage system. More importantly for our investigation of the ethnoscientific knowledge of the Kamilaroi and their neighbours, the clever men or elders who prepare for the Morning Star Ceremony have to understand the movements of the Morning Star/Venus well enough to predict when the Morning Star would rise for the first time, which is the time of the ceremony. Venus has an eight-year cycle of movement that requires careful observation and recording of the rising time to be able to predict a specific rising time and date for a ceremony. In the case of the first rising of the Morning Star, which is when the ceremony takes place, it occurs approximately every 584 days.

### 3.5.13 Meteors

Parker and Lang (1905: 69) reported that her Euahlayi sources said "if big meteor falls, followed by a thunderclap, sign that great man has died. Should a number of stars shoot off from falling star, sign that man has died leaving large family. When star seen falling in daytime, sign that one of the Noongahburra tribes has died". Parker and Lang (1905: 74) also say that a meteor always means death. The Kamilaroi have a description of meteors; *miri yanan* (literally, "star go"), but Peck (1925: 1-5) recorded stories about meteors from the "basalt country where the waratah does not grow", which could refer to the western ranges of the Dividing Range in Kamilaroi country (a waratah is a red flowering plant). In one story, a hunter got in a fight with an opposing tribe, after which people missing from his tribe became waratahs. "A great bright light, burning blue, travelling at enormous rate" wiped out the opposing tribe. The other story is about a "hot day, followed by a night when the heavens literally split up, and the star groups, loosened from their holds, came flashing to Earth, with millions of pieces of molten objects, leaving burning holes, which became waratahs." Mathews & White (1994: 60) speculate that a Murrawarri story of Gien, the Moon, who dropped a large sheet of bark on people who left him to drown (see Murrawarri story of the Moon), had left a large scar on the Earth near Boneda, NSW, which we speculate may be a meteor impact crater. The Wailwun had a story told by June Barker (in MacKay 2001: 112-4) about *Gambil Gambil*, a spirit woman who used to roam Ngemba and Murrawarri country. After she killed a number of the people, the Wailwun asked a clever man to do something about *Gambil Gambil*. He managed to track her down, and tied himself to her back with a hair cord. She flew into the sky to shake him off, where a falling star knocked him off, and the star and clever man fell near Girilambone, NSW, where it lit up the land around. *Gambil Gambil* is still flying around as a falling star.

June Barker, who identified as Kamilaroi (but told the Ngemba story above), told P6 the Kamilaroi name for shooting star (*miri yanan*), and said that "if they lost someone (someone died), the old people would sit up all night and wait for a shooting star to appear, to let them know that person had reached camp. A falling star appeared when a life is taken and a life given." P1, also Kamilaroi, repeated that "shooting star – person dies; same shooting star brings new baby". These stories confirm the Murrawarri belief, expressed by P5:

> When someone dies, and becomes a spirit, they go to the sky, and stay in the sky until they return to Earth on a falling star. When a baby is conceived, it only has the physical body, no spirit. When it is born, it gets a spirit, which comes back on a shooting star, and waits behind a *yarran* tree, which is one of the birthing areas.

This is a reference story to the importance of maintaining country, and was connected to the importance of the Gooraman Swamp, near Weilmorangle, NSW, which has many of these trees. We note the striking similarity of this story to that of the Wardaman people of the Northern Territory (Cairns and Harney, 2003).

It would seem that the Euahlayi are unlike the other groups in regards to how the birth spirit gets to Earth; for them it comes on the Moon's rays, but for the other people, it comes on a meteor.

Hamacher and Norris (2009) present a detailed geomythology survey of Aboriginal stories about falling stars, to link them to impact craters in Australia, but there are no confirmed impact craters in the area of this study, other than a large, but extremely old crater (millions of years BP) near White Cliffs, NSW, and another, unconfirmed smaller crater (Green, 2008; Hamacher and Norris 2010: 12).

Two of the participants (P2 and P7) have referred to the story of the broken-neck brush turkey (see Scorpius), and the meteor that broke its head. We were told that there is a ceremonial connection, but whether that is to the turkey or the meteor is unclear.

## 4. Discussion

### 4.1 Background of the stories

One of the major issues with Aboriginal cultural stories, mentioned in the Introduction as the controversy introduced by Swain, is whether or not the stories reflect pre-invasion culture, or have been significantly altered post-invasion. In other studies of Aboriginal cultural stories related to the sky, it has been suggested that rock art may be used to establish whether stories are pre-invasion. In this study, rock art could not be used to date stories because of cultural sensitivities, and so we needed another method of dating stories. The Reference Group advised that, in Kamilaroi/Euahlayi culture, young people were taught stories and knowledge of the sky by their grandfathers, so that oral knowledge was transmitted in such a way that it skipped a generation. Sveiby & Skuthorpe (2006: 53-6) also stated that this was earlier practice.

Ridley (1856) was the earliest writer to describe Kamilaroi culture. His later work, *Report on Australian Language and Traditions* (Ridley, 1873), had a significant section detailing names and descriptions of objects in the sky, based on an "evening under the sky" with an Aboriginal person identified as King Rory. Ridley further identified King Rory as "Ippai Dinoun, the Chief of the Gingi Tribe". The Gingi clan can be traced to what was later called Gingie Station, about five kilometres north of Walgett, making King Rory likely to have been a Euahlayi person. Ridley said that King Rory was about 60 years of age (in 1873). Assuming that King Rory was about 15 years old when his grandfather taught him the knowledge of the sky that he passed on to Ridley, this transmission of knowledge occurred around 1828, which is three years before the first white explorer, Mitchell, came anywhere near Walgett. Walgett itself wasn't occupied by non-Aboriginal people until the 1850's, and there is no record of any missionary activity in this area until later, so this instance of knowledge in the literature is strong evidence for a pre-invasion origin.

King Rory's stories are similar to those collected during the ethnographic part of this study. He described an emu in the Coalsack as *gao-ergi*, which is very close to the current Kamilaroi/Euahlayi *Gawarrgay*, and the Milky Way as *Worambul*, which is close to *Warrambul* (Ash et al 2006: 82, 139). The dating of the transmission of knowledge to King Rory, and the correlation with stories included in this study, strongly indicate that, contrary to Swain's arguments, they represent pre-invasion cultural knowledge of the Kamilaroi and their neighbours.

### 4.2 "What's up there is down here"

Several participants described an object or story in the sky as having a mirror existence on Earth, with the descriptions such as "what's up there is down here", and "the Milky Way represents where things are – campsites, tribes, ancestral places, in other words, a sky atlas". Examples include:

1. the head of the Emu in the Sky represents a waterhole on Earth, and the Emu looks after everything that lives there;

2. the Narran and Coocoran Lakes were formed by the crocodiles who took *Baayami's* wives, which are now the *garriyas* "crocodiles in the Milky Way";
3. ceremonial stories were shown in the sky, along with stories of culture heroes, such as *Baayami's* travels with his wives.

Many of the stories were related to physical things on the ground, such as the clever man chasing the *Miyaymiyaay*, and leaving rock pillars such as those found in the Warrumbungles, and the *bullis* in the Walgett/Brewarina area which were *Wilbaarr's* dark spots near Scorpius. One of the participants advised that everything up in the sky was once down on Earth because that's the way it started out, and the sky and the Earth reversed. This concept is significant in understanding why sky knowledge is so important in ceremony, and even everyday life.

### 4.3 Sky knowledge of the Kamilaroi and their neighbours

The outcome of the literature survey part of this study revealed a complex, but somewhat fragmented, sky knowledge mainly based on Kamilaroi vocabulary from Ridley, Euahlayi stories from Parker, and some ethnography from Mathews, with much of the information simply repeated by later authors. The collected stories in the ethnographic survey came from a cross-section of the Kamilaroi, Euahlayi, Ngemba, and Murrawarri communities, and to our surprise, there were some totally new stories that complemented the literature. Most importantly, the collected stories explained and enhanced the stories in the literature, which in many cases were repeated by different authors with parts either missing, or misinterpreted.

Section 3 provides at least one complete story from one language group about each of the 13 objects chosen, and in many cases there are different stories reflecting other language groups. Most importantly, some of the participants who have worked with this cultural information have provided something close to a linked cultural story of sky knowledge of the Kamilaroi and their neighbours. There are holes in the stories that were lost and may never be recovered, but given the conditions that Aboriginal people lived in after they were forced from much of their country, the resilience of their culture is impressive. We conclude that the Kamilaroi and their neighbours have retained a complex knowledge of the sky in their cultures that compares favourably with any other known sky culture in Australian Aboriginal studies to date.

We therefore conclude that our evidence supports our Hypothesis 1 that "knowledge from these language groups could add to the current body of knowledge of Aboriginal sky culture."

## 4.4 Ethnoscientific knowledge of the sky

This study has offered several instances of sky knowledge that are evidence for an ethnoscientific understanding of the sky. For example:

1. the Euahlayi (and possibly the Kamilaroi) explanation for solar eclipses is shared by two other Aboriginal language groups, the Wirangu and a group from northwest Arnhem Land. The explanation that the Sun is chasing the Moon, and overtakes it, is very close to ethnoscientific in description and shows an understanding that the two bodies are in the sky and that a solar eclipse is caused when the Moon passes in front of the Sun;

2. if the Euahlayi and Kamilaroi practice of the Morning Star ceremony requires exact knowledge of date and time of the rising of Venus before the sun, then this is a very strong argument for ethnoscientific knowledge, as Venus' eight year cycle of movement would require observation for at least this time to confirm a predicted rising for the ceremony. However, cultural sensitivity precludes us from discussing this in more detail;

3. a method for knowing the exact day of the year to predict the rising of Venus implies some form of calendar, which in itself requires an ethnoscientific approach.

Other examples which require further study were raised by a participant (P2) and include knowledge of the movement of objects in the sky so that stories can be told at the correct time of the object's visibility. Such putative examples of ethnoscientific knowledge of the sky by the Kamilaroi and their neighbours should be the subject of further research.

We therefore conclude that while we have only one firm piece of evidence (the understanding of eclipses) to support Hypothesis 2 that "the Kamilaroi and their neighbours had a ethnoscientific knowledge of the night sky through observation and experimentation", there exists significant circumstantial evidence that merits further study in this area.

## Acknowledgements

We acknowledge and pay our respects to the traditional owners and elders, both past and present, of the Kamilaroi, Euahlayi, Ngemba, and Murrawarri peoples. We thank the participants, Michael Anderson, Rhonda Ashby, Lachie Dennis, Paul Gordon, Greg Griffiths, Brenda McBride, Jason Wilson, and one anonymous person, for their stories and those of their families. We particularly thank Michael Anderson and Greg Griffiths for creating community interest, helping find other participants, and acting as our Reference Group.Wait, correcting - acknowledgements should be tagged:We acknowledge and pay our respects to the traditional owners and elders, both past and present, of the Kamilaroi, Euahlayi, Ngemba, and Murrawarri peoples. We thank the participants, Michael Anderson, Rhonda Ashby, Lachie Dennis, Paul Gordon, Greg Griffiths, Brenda McBride, Jason Wilson, and one anonymous person, for their stories and those of their families. We particularly thank Michael Anderson and Greg Griffiths for creating community interest, helping find other participants, and acting as our Reference Group.